\documentclass[a4paper,11pt]{article}
\usepackage{jcappub}
\usepackage[T1]{fontenc}
\title{\boldmath Gravitational Waves in Ghost Free Bimetric Gravity}

\author{Morteza Mohseni}

\affiliation{Physics Department, Payame Noor University, 19395-3697 Tehran, Iran}

\emailAdd{m-mohseni@pnu.ac.ir}

\abstract{We obtain a set of exact gravitational wave solutions for the ghost free bimetric theory of gravity. With a flat reference metric, the theory admits the vacuum Brinkmann plane wave
solution for suitable choices of the coefficients of different terms in the interaction potential. An exact gravitational wave solution corresponding to a massive scalar mode is also
admitted for arbitrary choice of the coefficients with the reference metric being proportional to the spacetime metric. The proportionality factor and the speed of the wave are calculated in
terms of the parameters of the theory. We also show that a $F(R)$ extension of the theory admits similar solutions but in general is plagued with ghost instabilities.
}

\begin{document}
\maketitle
\flushbottom

\section{Introduction}
Following the pioneering work by Fierz and Pauli \cite{fierz}, there have been numerous efforts to construct a consistent theory of massive gravitons,  see \cite{ruba} and \cite{hinter} for
reviews. Most of the resulted models suffered from subtle theoretical problems such as the van Dam-Veltman-Zakharov discontinuity \cite{van,velt} and the appearance of ghosts and
subsequent instability \cite{deser}, which make them far from being conclusive. Such efforts have recently been intensified, partly as a consequence of the belief that massive gravitons might
explain the observed accelerated expansion of the universe \cite{dub,eck,niz,gaba,amic,come,muk,cord,heis,volk,koy,lima}. In a remarkable breakthrough, a nonlinear extension of the Fierz-
Pauli model has been introduced recently \cite{rah1,rah2} which is shown to be free of the Boulware-Deser ghosts \cite{ros1}. Thus, the model is usually referred to as the ghost free massive
gravity. This model has attracted a lot of attentions, a nonexhaustive list of the relevant references includes
\cite{appl1,appl2,appl3,appl4,appl5,appl6,appl7,appl8,appl9,appl10,appl11,appl12}.

A particular feature of the massive gravity theories is the simultaneous existence of at least two metrics, the dynamical or full metric which specify the geometry of the spacetime, and the
absolute or reference metric which is related to the graviton mass term. In the ghost free model of massive gravity, the reference metric is fixed while the full metric is dynamical, in general.
On the other hand, in Ref. \cite{sch}, it has been shown that considering a dynamical reference metric does not destroy the ghost free property of this nonlinear theory of massive gravity. Thus,
it would be interesting to construct a model in which both metrics are dynamical. Actually, this has been done in Ref. \cite{hassan} in the context of  a bimetric theory. Basically, this is an
extension of the ghost free model of massive gravity in which the reference metric has its own kinetic term (an Einstein-Hilbert-like term) and an interaction term between the two metrics
with a form similar to the nonlinear potential of the ghost free model of massive gravity.\footnote{Actually, it seems that there is no standard terminology regarding these two types of metrics.
In Ref. \cite{viser}, the terms "background" and "foreground" are used for these metrics. Here, we reserve the term {\it reference} for the metric introduced to define the graviton mass and
the term {\it spacetime} for the one which determines the causal structure of spacetime.} Spherically symmetric solutions of the model have been studied in Ref. \cite{seper}. The
cosmological solutions to the model have been presented in Refs. \cite{volf} and \cite{straus}. A mechanism which removes the ghost from the model is discussed in Ref. \cite{hesam}. An
equivalent model has been formulated in terms of vielbeins in Ref. \cite{rosa}. A modified version of the model has been constructed \cite{odint} in which the curvature scalar of the spacetime 
metric is replaced with an arbitrary function of it using an the auxiliary scalar field and it was shown that the modified model admits a cosmological solution corresponding to an accelerating
universe. In Ref. \cite{paul} it was shown that the model can be extended by replacing the curvature scalars with the Lovelock invariants.

In this work we seek classes of exact pp-wave solutions of the ghost free bimetric gravity. The first motivation for this comes from the general interest on gravitational waves. In fact, 
gravitational waves are among the most important theoretical predictions of theories of gravity and are expected to be observed in near future. Thus, one expects any modified theory of gravity 
to accommodate such solutions. They are also of great interest in other areas of theoretical physics such as string theory, see, e.g., \cite{blau} and references therein. Here, we present a set of
massless and massive exact gravitational pp-wave solutions. This is particularly interesting because most of the known gravitational wave solutions to modified theories of gravity, models of 
massive gravity in particular, are obtained within the framework of linearized approximation, see e.g., \cite{capo} and references therein. For the case of massive gravity, which is not free of 
ghosts at the linear order, it would be more plausible to search for solutions describing strong gravitational waves as well. An Aichelburg-Sexl-type plane wave solution has been found for the 
ghost free massive gravity \cite{appl6} and a few other modified theories of gravity \cite{mohseni}. The second motivation lies in the bigravity model itself. Investigating new solutions to the 
model sheds more light on the physical content of the model and its potential strengths. The data associated with new solutions, including gravitational wave ones, might be used to constraint the 
model and its parameters.  

In the following sections we start with a review of the ghost free model of bimetric gravity and then present massless and massive pp-wave solutions in the absence of matter fields.
We show that the model admits vacuum plane gravitational solution for a suitable choice of the parameters by choosing a flat reference metric. For massive solutions to be admitted, we
can choose a reference metric proportional to the spacetime metric. We show that the speed of the wave can be calculated in terms of the parameters of the theory which at the same time
fix the proportionality factor. We also show that the $F(R)$ extension of the theory admits these solutions but it suffers from ghost instabilities. 

\section{The ghost free bigravity}
The action for the ghost free bimetric gravity \cite{hassan} may be written in the following form \cite{straus}
\begin{eqnarray}\label{e1}
S&=&-\frac{M^2_p}{2}\int{\sqrt{-g}}d^4xR+\int\sqrt{-g} d^4x{\mathcal L}_m(g,\Phi)
\nonumber\\&&-\frac{{\mathcal M}^2_p}{2}\int\sqrt{-f}d^4x{\mathcal R}+m^2M^2_p\int\sqrt{-g}d^4xU(X)
\end{eqnarray}
where $R$ and ${\mathcal R}$ are the curvature scalars associated with the spacetime metric $g_{\mu\nu}$ and the reference metric $f_{\mu\nu}$ respectively, ${\mathcal L}_m$ is the
Lagrangian density for the matter fields generically denoted by $\Phi$, $X=\sqrt{g^{-1}f}$, and
\begin{eqnarray}\label{e2}
U(X)=\sum_{n=0}^4\beta_n e_n(X)
\end{eqnarray}
 is the interaction potential for the two metrics $g_{\mu\nu}$ and $f_{\mu\nu}$. Here, $\beta_n$ are arbitrary constants and
\begin{eqnarray}\label{e3}
e_0&=&1,\nonumber\\
e_1&=&[X],\nonumber\\
e_2&=&\frac{1}{2}\left([X]^2-[X^2]\right),\\
e_3&=&\frac{1}{6}\left([X]^3-3[X][X^2]+2[X^3]\right),\nonumber\\
e_4&=&det(X)\nonumber
\end{eqnarray}
where square brackets stand for the trace with respect to the spacetime metric $g_{\mu\nu}$. Thus, the parameters of the model are the five constants $\beta_n$,
the coupling constants $M_p$ and ${\mathcal M}_p$, and the mass $m$. Variation of the above action with respect to $g_{\mu\nu}$ and $f_{\mu\nu}$
results in the following equations of motions, respectively
\begin{eqnarray}
G_{\mu\nu}&=&-\frac{m^2}{2}\sum_{n=0}^3(-1)^n\beta_n\left(g_{\mu\lambda}Y^\lambda_{(n)\nu}(X)+g_{\nu\lambda}Y^\lambda_{(n)\mu}(X)\right)+\frac{1}{M^2_p}T_{\mu\nu},
\label{e4}\\
{\mathcal G}_{\mu\nu}&=&-\frac{m^2M^2_p}{2{\mathcal M}^2_p}\sum_{n=0}^3(-1)^n\beta_{4-n}\left(f_{\mu\lambda}Y^\lambda_{(n)\nu}
(X^\prime)+f_{\nu\lambda}Y^\lambda_{(n)\mu}
(X^\prime)\right),\label{e5}
\end{eqnarray}
where $G_{\mu\nu}$ is the Einstein tensor for $g_{\mu\nu}$,  ${\mathcal G}_{\mu\nu}$ is the Einstein tensor associated with $f_{\mu\nu}$, $T_{\mu\nu}$ represents the energy-momentum 
tensor, $X^\prime=\sqrt{f^{-1}g}$, and $Y_{(n)}(X)$ are given by
\begin{eqnarray}\label{e6}
Y_{(0)}(X)&=&1,\nonumber\\
Y_{(1)}(X)&=&X-[X],\nonumber\\
Y_{(2)}(X)&=&X^2-X[X]+\frac{1}{2}\left([X]^2-[X^2]\right),\\
Y_{(3)}(X)&=&X^3-X^2[X]+\frac{1}{2}X\left([X]^2-[X^2]\right)-\frac{1}{6}\left([X]^3-3[X][X^2]+2[X^3]\right).\nonumber
\end{eqnarray}
\section{Massless gravitational waves}
In this section we look for massless plane gravitational wave solutions propagating in a flat background. Thus, adopting the chart $(u,v,x,y)$ in which $u=t-z$ and $v=t+z$ are the null
coordinates, we set
\begin{equation}\label{e7}
g_{\mu\nu}=\left(
\begin{array}{cccc}
0 & -\frac{1}{2} & 0 & 0\\
-\frac{1}{2} & 0 & 0 & 0\\
0 & 0 & 1 & 0\\
0 & 0 & 0 &1
\end{array}\right)
\end{equation}
which is the Minkowski metric, and
\begin{equation}\label{e8}
f_{\mu\nu}=\left(
\begin{array}{cccc}
-H & -\frac{1}{2} & 0 & 0\\
-\frac{1}{2} & 0 & 0 & 0\\
0 & 0 & 1 & 0\\
0 & 0 & 0 &1
\end{array}\right)
\end{equation}
where $H=H(u,x,y)$ is the wave profile. This corresponds to the following line element
\begin{equation}\label{e9}
ds^2=-dudv-H(u,x,y)du^2+dx^2+dy^2
\end{equation}
which is the Brinkmann expression for the plane wave. This results in
\begin{equation}\label{e8a}
{X^\mu}_\nu={X^\prime}^\mu_\nu=\pm\left(
\begin{array}{cccc}
1 & 0 & 0 & 0\\
H & 1 & 0 & 0\\
0 & 0 & 1 & 0\\
0 & 0 & 0 & 1
\end{array}\right)
\end{equation}
For the positive roots, inserting these into Eq. (\ref{e4}) results in
\begin{eqnarray*}
0=\beta_0\left(
\begin{array}{cccc}
0 & -1 & 0 & 0\\
-1 & 0 & 0 & 0\\
0 & 0 & 2 & 0\\
0 & 0 & 0 & 2
\end{array}\right)-
\beta_1\left(
\begin{array}{cccc}
-H & 3 & 0 & 0\\
3 & 0 & 0 & 0\\
0 & 0 & -6 & 0\\
0 & 0 & 0 & -6
\end{array}\right)+
\beta_2\left(
\begin{array}{cccc}
2H & -3 & 0 & 0\\
-3 & 0 & 0 & 0\\
0 & 0 & 6 & 0\\
0 & 0 & 0 & 6
\end{array}\right)-
\beta_3\left(
\begin{array}{cccc}
-H & 1 & 0 & 0\\
1 & 0 & 0 & 0\\
0 & 0 & -2 & 0\\
0 & 0 & 0 & -2
\end{array}\right)
\end{eqnarray*}
from which we obtain
\begin{eqnarray}\label{e10}
\beta_2&=&-\beta_0-2\beta_1,\nonumber\\
\beta_3&=&2\beta_0+3\beta_1.
\end{eqnarray}
Similarly, the other equation of motion, Eq. (\ref{e5}), reduces to
\begin{eqnarray*}
\left(
\begin{array}{cccc}
-\frac{1}{2}\nabla^2H  & 0 & 0 & 0\\
0 & 0 & 0 & 0\\
0 & 0 & 0 & 0\\
0 & 0 & 0 & 0
\end{array}\right)+
\alpha_4\left(
\begin{array}{cccc}
-2H & -1 & 0 & 0\\
-1 & 0 & 0 & 0\\
0 & 0 & 2 & 0\\
0 & 0 & 0 & 2
\end{array}\right)-
\alpha_3\left(
\begin{array}{cccc}
5H & 3 & 0 & 0\\
3 & 0 & 0 & 0\\
0 & 0 & -6 & 0\\
0 & 0 & 0 & -6
\end{array}\right)\\+
\alpha_2\left(
\begin{array}{cccc}
-4H & -3 & 0 & 0\\
-3 & 0 & 0 & 0\\
0 & 0 & 6 & 0\\
0 & 0 & 0 & 6
\end{array}\right)-
\alpha_1\left(
\begin{array}{cccc}
H & 1 & 0 & 0\\
1 & 0 & 0 & 0\\
0 & 0 & -2 & 0\\
0 & 0 & 0 & -2
\end{array}\right)=0
\end{eqnarray*}
in which $\alpha_n=\frac{m^2M^2_p}{2{\mathcal M}^2_p}\beta_n$, and $\nabla^2$ is the Laplacian in the transverse plane. Taking Eq. (\ref{e10}) into account, this results in
\begin{equation}\label{e11}
\beta_4=-3\beta_0-4\beta_1,
\end{equation}
and
\begin{equation}\label{e12}
\nabla^2 H=-\frac{m^2M^2_p}{{\mathcal M}^2_p}(2\beta_4+5\beta_3+4\beta_2+\beta_1)H
\end{equation}
which upon substitution of the above obtained values of $\beta_{2,3,4}$ reduces to
\begin{equation}\label{e13}
\nabla^2 H=0
\end{equation}
which is the standard equation for the wave profile in general relativity. For the negative roots in Eq. (\ref{e8a}) we obtain the same results but with $\beta_1\rightarrow-\beta_1$ and
$\beta_3\rightarrow-\beta_3$.

Here, two more points are in order. First, because in the absence of matter fields the action is invariant under the following interchanges \cite{straus}
\begin{equation}\label{e14}
g_{\mu\nu}\leftrightarrow f_{\mu\nu},\hspace{2mm} \beta_n\leftrightarrow\beta_{4-n},\hspace{2mm}M_p\leftrightarrow{\mathcal M}_p,
\hspace{2mm}m^2\rightarrow\frac{m^2M^2_p}{{\mathcal M}^2_p},
\end{equation}
we have automatically another solution in which $f_{\mu\nu}$ is equal to the Minkowski metric and $g_{\mu}$ is the plane wave. Second, if we generalize the model by replacing ${\mathcal
R}$ in Eq. (\ref{e1}) with an arbitrary function of  it, i.e., ${\mathcal F}(\mathcal R)$, then the first term in the left-hand side of Eq. (\ref{e5}) gets modified as
\begin{eqnarray*}
{\frac{d{\mathcal F(\mathcal R)}}{d{\mathcal R}}}{\mathcal R}_{\mu\nu}-\frac{1}{2}{\mathcal F}({\mathcal 
R})f_{\mu\nu}+\left(f_{\mu\nu}\square_{(f)}-\nabla_{(f)\mu}\nabla_{(f)\nu}\right)\frac{d{\mathcal F}(\mathcal R)}{d\mathcal R}
\end{eqnarray*}
while the other terms (and also the other equation of motion, Eq. (\ref{e4})) remains intact. It can easily be shown that it also admits the same solution as presented above provided
${\mathcal F}(0)=0$. In the absence of matter fields, the same thing is also true for the case where the curvature scalar of the spacetime metric is promoted to $F(R)$, as in Ref. \cite{odint}.  
It should be noted that in general, the extra degrees of freedom associated with these extensions may lead to appearance of ghosts in the model. In 
some specific situations, these ghosts can be avoided \cite{odint}.
\section{Massive gravitational waves }
In this section we seek massive pp-wave solutions in vacuum. We switch to the Cartesian coordinates $(t,x,y,z)$ and start with the following ansatz
\begin{eqnarray}
g_{\mu\nu}&=&K(\beta t-z)\eta_{\mu\nu},\label{e15}\\
f_{\mu\nu}&=&\alpha^2 K(\beta t-z)\eta_{\mu\nu}\label{e16}
\end{eqnarray}
in which $0<\beta<1$ and $\alpha$ are constants, and the specific form of the function $K(\beta t-z)$ is to be determined from the equations of motion. With this ansatz, we have
\begin{equation}\label{e17}
X^\mu_\nu=\pm \alpha\delta^\mu_\nu
\end{equation}
and
\begin{equation}\label{e17a}
{X^\prime}^\mu_\nu=\pm \alpha^{-1}\delta^\mu_\nu
\end{equation}
from which we obtain (for the positive signs)
\begin{eqnarray}
\sum_{n=0}^3(-1)^n\beta_n(g_{\mu\lambda}Y^\lambda_{(n)\nu}(X)+g_{\nu\lambda}Y^\lambda_{(n)\mu}(X))&=&2(\beta_0+3\alpha\beta_1+3\alpha^2\beta_2
\nonumber\\&&+\alpha^3\beta_3)K(\beta t-z)\eta_{\mu\nu},\label{e21}\\
\sum_{n=0}^3(-1)^n\beta_n(f_{\mu\lambda}Y^\lambda_{(n)\nu}(X^\prime)+f_{\nu\lambda}Y^\lambda_{(n)\mu}(X^\prime))&=&2(\alpha^2\beta_4+3\alpha\beta_3+3\beta_2
\nonumber\\&&+\alpha^{-1}\beta_1)K(\beta t-z)\eta_{\mu\nu}.\label{e21a}
\end{eqnarray}
On the other hand, the only nonvanishing off-diagonal component of the Einstein tensor associated with the metric given in Eq. (\ref{e15}) is
\begin{equation}\label{e22}
G_{tz}=\frac{\beta}{2}\frac{2K(\beta t-z)K^{\prime\prime}(\beta t-z)-3(K^\prime(\beta t-z))^2}{(K(\beta t-z))^2}
\end{equation}
where primes denote differentiation with respect to $\beta t-z$. Now, Eq. (\ref{e4}) requires $G_{tz}$ to vanish and hence
\begin{equation}
K(\beta t-z)=K_0(\beta t-z)^{-2},\label{e23}\\
\end{equation}
with $K_0$ being a constant. Now, the Einstein tensor reduces to
\begin{equation}\label{e26}
G_{\mu\nu}=-3(1-\beta^2)(\beta t-z)^{-2}\eta_{\mu\nu},
\end{equation}
and a similar relation for ${\mathcal G}_{\mu\nu}$. Inserting these into Eq. (\ref{e4}) we obtain
\begin{equation}\label{e27}
1-\beta^2=\frac{m^2}{3}(\beta_0+3\alpha\beta_1+3\alpha^2\beta_2+\alpha^3\beta_3)K_0.
\end{equation}
Similarly, from Eq. (\ref{e5}) we obtain
\begin{equation}\label{e27a}
1-\beta^2=\frac{m^2 M^2_p}{3{\mathcal M}^2_p}(\alpha^2\beta_4+3\alpha\beta_3+3\beta_2+\alpha^{-1}\beta_1)K_0.
\end{equation}
Thus, the model admits the solution given in Eqs. (\ref{e15}) and (\ref{e16}) provided the above two conditions holds. This corresponds to a scalar massive wave propagating in the $z$-
direction with subluminal speed $\beta$. The above two equations determine the values of $\beta$ and $\alpha$ in terms of the parameters of the model. As an example, by choosing
$\beta_2$ to be the only nonvanishing $\beta_n$, one obtains $\alpha=\frac{M_p}{{\mathcal M}_p}$ and $\beta=\sqrt{1-\frac{m^2M^2_p}{{\mathcal M}^2_p}K_0\beta_2}$. Then the
constraint $\beta<1$ results in $0<K_0\beta_2<\frac{{\mathcal M}^2_p}{m^2M^2_p}$. Starting with the negative roots in Eqs. (\ref{e17}) and (\ref{e17a}) results in similar relations but 
with $\beta_1\rightarrow-\beta_1$ and $\beta_3\rightarrow -\beta_3$.

\section{The $F(R)$ extension}
For the extended model where in action (\ref{e1}) the curvature scalar $R$ is replaced with a generic function $F(R)$,  we obtain, instead of Eq. (\ref{e4}), the following equation of motion
\begin{eqnarray}
(R_{\mu\nu}+g_{\mu\nu}\square-\nabla_\mu\nabla_\nu)F^\prime(R)-\frac{1}{2}g_{\mu\nu}F(R)&=&-\frac{m^2}
{2}\sum_{n=0}^3(-1)^n\beta_n\left(g_{\mu\lambda}Y^\lambda_{(n)\nu}(X)+\mu\leftrightarrow\nu\right)\nonumber,
\\ \label{e4a}
\end{eqnarray}
while the other equation of motion, Eq. (\ref{e5}) remains unchanged. Here, $F^\prime(R)=\frac{dF(R)}{dR}$ and we have assumed that matter fields are absent. Now, repeating the 
calculations of the previous section, one can show that the massive pp-wave solution given in Eqs. (\ref{e15}) and (\ref{e16}) is still admitted provided Eq. (\ref{e27}) is replaced with
\begin{equation}\label{e27c}
1-\beta^2=\frac{m^2}{3F^\prime(R_0)}\left(\beta_0+3\alpha\beta_1+3\alpha^2\beta_2+\alpha^3\beta_3+\frac{F(R_0)}{2m^2}\right)K_0,
\end{equation}
where the curvature scalar $R_0=12(1-\beta^2)K^{-1}_0$ is constant. The consistency of the equations constraints the function $F(R)$. For the example case where $\beta_{0,1,3,4}=0, \beta_2\neq 0$,  we obtain $\alpha=\sqrt{F^\prime(R_0)\left(\frac{M^2_p}{{\mathcal M}^2_p}-\frac{F(R_0)}{6m^2F^\prime(R_0)\beta_2}\right)}$ which requires that the relation $\frac{F(R_0)}{F^\prime(R_0)}<\frac{6M^2_pm^2\beta_2}{{\mathcal M}^2_p}$ holds.

The massive wave solution is a constant curvature spacetime which can be used to explore the properties of the $F(R)$ extension further. This can be achieved by considering the propagating 
modes resulted from the extension. Taking the trace in both sides of Eq. (\ref{e4a}), we obtain
\begin{equation}\label{e28}
3\square F^\prime(R)+RF^\prime(R)-2F(R)=\Delta
\end{equation}
where $\Delta\equiv -m^2\sum_{n=0}^3(-1)^n\beta_n Y^\mu_{(n)\mu}(X)$. Now, defining $\phi\equiv F^\prime(R)$ and $\frac{dV}{d\phi}=\frac{1}{3}(2F(R)-RF^\prime(R)+\Delta)$,
we reach the following Klein-Gordon equation
\begin{equation}\label{e29}
\square\phi=\frac{dV}{d\phi}
\end{equation}
in which $V$ plays the role of an effective potential. For the spacetime under consideration, the minimum of $V$ occurs at $\phi_0$ where the following relation is satisfied
\begin{equation}
2F(R_0)-R_0F^\prime(R_0)+\Delta=0.
\end{equation} 
For the perturbation $\phi=\phi_0+\delta\phi$ around the minimum value, we have
\begin{equation}\label{e30}
(\square-M^2_{FR})\delta\phi=0
\end{equation}
where $M^2_{FR}=\frac{1}{3}\left(\frac{F^\prime(R_0)}{F^{\prime\prime}(R_0)}-R_0\right)$. Since one expects $F^{\prime\prime}(R_0)$ to be very small \cite{faraoni}, 
$M^2_{FR}$ is positive. This gives rise to a massive propagating mode essentially similar to the $F(R)$ extension of massless gravity \cite{revi1}. However, the $F(R)$ extension
also excites ghost modes in the model. To see this more clearly, we first rewrite the action in terms of $g_{\mu\nu}=K(\beta t-z)\eta_{\mu\nu}$ and $f_{\mu\nu}=L(\beta t-z)\eta_{\mu\nu}$ with the assumption that matter fields are
absent. This reads $S=\int{\mathcal S}d^4x$ where
\begin{eqnarray}
{\mathcal S}&=&-\frac{M^2_p}{2}\left(\frac{3}{2}(1-\beta^2)\right)\left(2K^{\prime\prime}-\frac{{K^\prime}^2}{K}\right)-\frac{{\mathcal M}^2_p}{2}\left(\frac{3}{2}
(1-\beta^2)\right)\left(2L^{\prime\prime}-\frac{{L^\prime}^2}{L}\right)\nonumber\\&&+m^2M^2_p\left(\beta_0K^2+4\beta_1K{\sqrt {KL}}+6\beta_2KL+4\beta_3L{\sqrt 
{KL}}+\beta_4L^2\right)\label{e31}
\end{eqnarray}  
where primes stand for differentiation with respect to $u\equiv\beta t-z$. The associated equation of motion of $K(\beta t-z)$ can be obtained from the Euler-Lagrange equation
\begin{equation}\label{e32}
\sum_{n=0}(-1)^n\frac{d^n}{du^n}\left(\frac{\partial{\mathcal S}}{\partial\left(\frac{d^nK}{du^n}\right)}\right)=0
\end{equation}
which gives 
\begin{eqnarray}
\frac{3}{4}(1-\beta^2)\left(\frac{2K^{\prime\prime}}{K^2}-\frac{{K^\prime}^2}{K^3}\right)=2m^2\left(\beta_0+3\beta_1\sqrt{\frac{L}{K}}+3\beta_2\frac{L}{K}+\beta_3\frac{L}
{K}\sqrt{\frac{L}{K}}\right)K,\label{e33}
\end{eqnarray}
and a similar equation for $L(\beta t-z)$. These equation have consistent solutions $K(\beta t-z)=K_0(\beta t-z)^{-2}$ and $L(\beta t-z)=\alpha^2 K(\beta t-z)$ which upon insertion in Eq. 
(\ref{e33}) results in Eqs. (\ref{e27}) and (\ref{e27a}). 

For the extended model in which the scalar curvature $R$ (in the spacetime sector or the reference sector or both) is replaced with a generic function $F(R)$, the relevant kinetic terms in the 
right-hand side of Eq. (\ref{e31}) should be modified accordingly. For the case of the spacetime sector, the first term on the right-hand side of Eq. (\ref{e31}) should be replaced with
\begin{eqnarray*}
-\frac{M^2_p}{2}K^2F\left(\frac{3}{2}(1-\beta^2)\left[\frac{2K^{\prime\prime}}{K^2}-\frac{{K^\prime}^2}{K^3}\right]\right).
\end{eqnarray*}
With this, the first two terms in the left-hand side of Eq. (\ref{e32}) are as follows
\begin{eqnarray*}
\frac{d^2}{du^2}\left(\frac{dF(R)}{dR}\right)-\frac{d}{du}\left(-\frac{dF(R)}{dR}\frac{K^\prime}{K}\right)&=&\frac{d^2F(R)}
{dR^2}R^{\prime\prime}+\frac{d^3F(R)}{dR^3}(R^\prime)^2\\&&+\frac{K^\prime}{K}\frac{d^2F(R)}{dR^2}R^\prime+\left(\frac{K^{\prime\prime}}{K}
-\frac{{K^\prime}^2}{K^2}\right)\frac{dF(R)}{dR}
\end{eqnarray*} 
where we have dropped overall factors $-\frac{M^2_p}{2}$ and $3(1-\beta^2)$. In general, noting the relation $R=\frac{2K^{\prime\prime}}{K^2}-\frac{{K^\prime}^2}{K^3}$, the first three terms in the right-hand side of the above relation contain $K^{\prime\prime\prime\prime}$ and
$K^{\prime\prime\prime}$. Thus, the equation of motion for $K$ will be of the following general form
\begin{equation}\label{e48}
\frac{d^4K(u)}{du^4}+P\frac{d^3K(u)}{du^3}+Q\frac{d^2K(u)}{du^2}+\cdots=0.
\end{equation}
Thus, by considering a perturbed solution $$K=K^{(0)}+\delta K$$ where $K^{(0)}=K_0(\beta t-z)^{-2}$, we will end up with an equation of motion for $\delta K$ which is generally of fourth 
order derivatives. According to the Ostrogradski theorem, such equations result in ghost propagating modes \cite{revi2}. Therefore, the $F(R)$ extended model is plagued with ghost 
instabilities. Similar arguments hold when the extension is made in the reference sector.
\section{Conclusions}
We showed that the ghost free bimetric theory of gravity admits exact gravitational pp-wave solutions. For vacuum, the Brinkmann gravitational wave solution is admitted with a
flat reference metric for suitable choices of the parameters. The resulting constraints on the coefficients of different terms in the potential determines three of them in terms of the remaining
ones, namely, those of the zero and first order terms. We showed that a $F(R)$ modified version of the theory admits the same solution. An exact massive gravitational wave
solution is also admitted with a reference metric being proportional to the spacetime metric. The proportionality constant and the speed of the wave were calculated in terms of parameters of
the model. The massive solution corresponds to a scalar mode of gravitational wave. We showed that the $F(R)$ extension of the model also admits the solution but the massive solution is not 
stable because of the presence of ghost modes. The exact gravitational wave solutions presented here are useful for studying the strong gravitational
waves in the context of ghost free bimetric gravities and are more compatible with the nonlinear nature of such theories.

\acknowledgments
I would like to thank S.D. Odintsov for comments. Comments from an anonymous referee of JCAP helped improve the presentation of the manuscript. I acknowledge the Abdus Salam ICTP 
where part of this work was done.

\end{document}